\RequirePackage{lineno}
\documentclass[a4paper,11pt]{article}
\usepackage{pos}

\usepackage{graphicx} 
\usepackage{epstopdf}
\usepackage{overpic}
\graphicspath{{./figures/}}
\newsavebox{\tablebox}
\usepackage{comment}
\usepackage{booktabs}

\input{belle2sym}
\def\DzToKPiEta {D^{0}\to K^{-}\pi^{+}\eta}
\def\DzToPiPiEta {D^{0}\to\pi^{+}\pi^{-}\eta}
\def\DzToKKEta {D^{0}\to K^{+}K^{-}\eta}
\def\DzToPhiEta {D^{0}\to\phi\eta}

\title{Recent charm results at Belle}

\author*{Longke Li}

\affiliation{University of Cincinnati,\\
  Cincinnati, OH 45221, USA}

\emailAdd{lilk@ucmail.uc.edu}

\abstract{
We present the selected recent charm results at Belle: 
(1) the $\CP$ asymmetries and branching fractions for the decays of charm mesons, i.e. $\DzToPiPiEta,\,\Kp\Km\eta$ and $\DzToPhiEta$; $\Dsp\to\Kp\piz,\,\Kp\eta$ and $\Dsp\to\pip\piz,\,\pip\eta$. 
(2) the branching fractions for the decays of charmed baryons, i.e. $\Lcp\to p\piz,\,p\eta$; $\Lcp\to\Lambda\pip\eta,\,\Sigma^0\pip\eta$ and intermediate processes $\Lcp\to\Lambda(1670)\pip,\,\eta\Sigma(1385)^+$;  $\Xi_c^0\to\Lambda\Kstarzb,\,\Sigma^0\Kstarzb$ and $\Sigma^+\Kstarm$. 
(3) properties of excited charmed baryons, i.e. search for electromagnetic transition $\Xi_c(2790,2815)^{+,0}\to\Xi_c^{+,0}\gamma$; and determination of spin-parity of $\Xi_c(2970)^+$.
}

\FullConference{%
  *** 10th International Workshop on Charm Physics (CHARM2020), ***\\
  *** 31 May - 4 June, 2021 ***\\
  *** Mexico City, Mexico - Online ***
}

\begin{document}

\maketitle

\section{Belle experiment}
The Belle experiment ran at the 
KEKB energy-asymmetric collider~\cite{bib:KEKB}.
The only detector operating at KEKB, Belle detector, is a large-solid-angle magnetic spectrometer 
consisting of a silicon vertex detector, a $50$-layer central drift chamber, an array of aerogel threshold Cherenkov counters, a barrel-like arrangement of time-of-flight scintillation counters, and an electromagnetic calorimeter comprising CsI(Tl) crystals located inside a superconducting solenoid coil providing a $1.5$T magnetic field. 
A detailed description of the 
detector is given in Refs.~\cite{bib:BelleDetector}. 
These subdetectors contribute to Belle to be a detector with advanced performances on the momentum and vertex resolution; the particle identification; and so on.
Belle accumulated her final data set of integrated luminosity 1 $\invab$~\cite{bib:BelleDetector} more than ten years ago. This data set provides us a large charm sample to study charm physics. In this proceedings, we present some selected recent charm results at Belle.

\section{$\CP$ asymmetries and Branching fractions for the decays of charm mesons}
The time-integrated $\CP$ asymmetry is defined as 
$\Acp=(\mathcal{B}(D\to f) - \mathcal{B}(\Db\to\fbar))/(\mathcal{B}(D\to f) + \mathcal{B}(\Db\to\fbar))$ where $\mathcal{B}$ is the branching fraction for $D\to f$ decays. Taking $D^0$ from $D^{*+}\to D^0\pip_s$ for example, it is produced from $e^{+}e^{-}\to c\bar{c}\to D^{*+}+X$ inclusive process, where the charge of $\pip_s$ is used to tag the $D^0$ flavor. 
The raw asymmetry is determined by the asymmetry of signal yields. 
It includes several sources for asymmetry: the $\CP$ asymmetry in $D^0$ decay, the final state detection asymmetry, and $D^*$ production forward-backward asymmetry $\Afb$, which is arising from the $\gamma$-$Z^0$ interference and higher-order QED effects.
Experimentally, we usually use two methods to remove these additional asymmetries: (1) use a reference mode of which $\Acp$ has been well-measured.
(2) perform a correction for the charged track detection asymmetry, e.g. weighting $\Dz$ and $\Dzb$ samples separately with the factors to correct the $\pip_s$ asymmetry: $w=1\mp A_{\varepsilon}^{\pi_s^+}[\cos\theta, p_T]$
where $A_{\varepsilon}^{\pi_s^+}$ is the $\pi_s^+$ efficiency map dependent on its polar angle and transverse momentum, and determined with the tagged and untagged $D^0\to\Km\pip$ samples. 
Then, the weighted samples have the corrected raw asymmetry, which only has two terms: $A_{\rm corr}=\Acp+\Afb$. 
Since $\Acp$ is independent on $\cos\theta^{*}$ and $\Afb$ is odd function of $\cos\theta^{*}$, we can obtain $\Acp$ and $\Afb$ in symmetric bins of $\cos\theta^{*}$:
\begin{eqnarray}
\Acp  = \dfrac{A_{\rm corr}(\cos\theta^{*})+A_{\rm corr}(-\cos\theta^{*})}{2}\,, \quad \quad 
\Afb = \dfrac{A_{\rm corr}(\cos\theta^{*})-A_{\rm corr}(-\cos\theta^{*})}{2}\,. \label{eqn:Acp}
\end{eqnarray}
Finally, fitting these $\Acp$ values to a constant gives the final measurement of $\Acp$ of our signal decay that we are interested in.

\subsection{$\Acp$ and $\mathcal{B}$ for $\Dz\to\pip\pim\eta$, $\Dz\to\Kp\Km\eta$, and $\Dz\to\phi\eta$}
Recently we present a measurement of $\CP$ asymmetries and branching fractions for three singly Cabibbo-suppressed (SCS) decays $\Dz\to\pip\pim\eta$, $\Kp\Km\eta$, and $\phi\eta$~\cite{Belle:2021dfa}.
All their current world averaged branching fractions have large uncertainties, especially $D^0\to\Kp\Km\eta$, of which significance is still below $5\sigma$. 
The Cabibbo-favored (CF) decay $D^0\to\Km\pip\eta$ is chosen as the reference mode, which has been well-measured (with a fractional uncertainty 
$\delta\mathcal{B}/\mathcal{B}\sim3\%$~\cite{bib:PDG2021}) by both Belle~\cite{bib:PRD102d012002} and BESIII~\cite{bib:PRL124d241803}.

Based on full Belle data set of 980 $\invfb$, we extract signal yields with fitting on $Q=M(h^+h^-\eta\,\pi_s)-M(h^+h^-\eta)-m_{\pi_s}$, which is the
kinetic energy released in the $D^{*+}$ decay and divided by $c^2$.
We find 13 thousands $\DzToPiPiEta$, and 1.5 thousands $\DzToKKEta$, and 660 yields for $\phi(1020)$-excluded $\DzToKKEta$, along with 180 thousands $\DzToKPiEta$.

Then we determine the efficiency-corrected yield across on Dalitz-plot to consider the bin-to-bin variation of efficiencies. This efficiency correction is based on the efficiency plane from PHSP signal MC, Dalitz-plot in $Q$ signal region, and in $Q$ sideband for background shape. 
Thus, we have the relative branching ratio by the efficiency-corrected yield ratio.
\begin{eqnarray}
\mathcal{B}(\DzToPiPiEta)/\mathcal{B}(\DzToKPiEta) & = & [6.49 \pm 0.09\,({\rm stat}) \pm 0.12\,({\rm syst})]\times10^{-2}\,, \\
\mathcal{B}(\DzToKKEta)/\mathcal{B}(\DzToKPiEta) & = & [9.57^{+0.36}_{-0.33}\,({\rm stat}) \pm 0.20\,({\rm syst})]\times10^{-3}\,, \\
\mathcal{B}(\DzToKKEta)_{{\rm ex.-}\phi}/\mathcal{B}(\DzToKPiEta) & = & [5.26\,^{+0.45}_{-0.38}\,(\rm{stat})\pm 0.11\,({\rm syst})]\times10^{-3}\,.
\end{eqnarray}
Using $\mathcal{B}(\DzToKPiEta)=(1.88\pm 0.05)\%$~\cite{bib:PDG2021}, we have the absolute branching fractions:
\begin{eqnarray}
\mathcal{B}(\DzToPiPiEta) & = & [1.22\pm 0.02\,({\rm stat}) \pm 0.02\,({\rm syst}) \pm 0.02\,(\mathcal{B}^{}_{\rm ref})]\times 10^{-3}\,, \\
\mathcal{B}(\DzToKKEta) & = &  [1.80\,^{+0.07}_{-0.06}\,({\rm stat}) \pm 0.04\,({\rm syst}) \pm 0.03\,(\mathcal{B}^{}_{\rm ref})] \times 10^{-4}\,, \\
\mathcal{B}(\DzToKKEta)_{{\rm ex.-}\phi} & = & [0.99\,^{+0.08}_{-0.07}\,({\rm stat})\pm 0.02\,({\rm syst})\pm  0.02\,(\mathcal{B}_{\rm ref})]\times 10^{-4}\,.
\end{eqnarray}
The last one is somewhat higher, but more precise, than a similar measurement by BESIII~\cite{bib:PRL124d241803}. 

In the Dalitz plot of $D^0\to\Kp\Km\eta$, we observe a very clear structure for $\phi(1020)$. Thus, we measure the branching fraction of this SCS and color-suppress decay $D^0\to\phi(1020)\eta$. 
To extract the signal yield, we perform $M_{KK}$-$Q$ two-dimensional fit instead of $Q$ one-dimensional fit, since there is the $Q$-peaking background from non-$\phi$ $D^0\to\Kp\Km\eta$ component. 
Finally, we obtain about 700 yields. The difference in likelihood, with and without including a signal component, is $\Delta\ln L=464.8$, which corresponds to a very high statistical significance ($>30\sigma$). This indicates we achieve the first observation of $D^0\to\phi\eta$. 
We determine the relative branching fraction:
\begin{eqnarray}
\mathcal{B}(D^0\to\phi\eta,\,\phi\to\Kp\Km)/\mathcal{B}(\Dz\to\Km\pip\eta)  = [4.82\pm 0.23\,({\rm stat})\pm 0.16\,({\rm syst})] \times 10^{-3}\,.
\end{eqnarray}
Using $\mathcal{B}(\DzToKPiEta)=(1.88\pm 0.05)\%$~\cite{bib:PDG2021} and $\mathcal{B}(\phi\to\Kp\Km)=(49.2\pm0.5)\%$~\cite{bib:PDG2021}, we have the absolute branching fraction, 
\begin{eqnarray}
\mathcal{B}(D^0\to\phi\eta) = [1.84\pm 0.09\,({\rm stat})\pm 0.06\,({\rm syst})\pm 0.05\,(\mathcal{B}_{\rm ref})]\times 10^{-4},
\end{eqnarray}
where the systematic uncertainty includes the small uncertainty on $\mathcal{B}(\phi\to\Kp\Km)$.
This result is consistent with the current world average of $(1.8\pm0.5)\times10^{-4}$~\cite{bib:PDG2021}, but notably more precise than it.

\begin{figure}[!htbp]
  \begin{center}
  \begin{overpic}[width=0.333\textwidth]{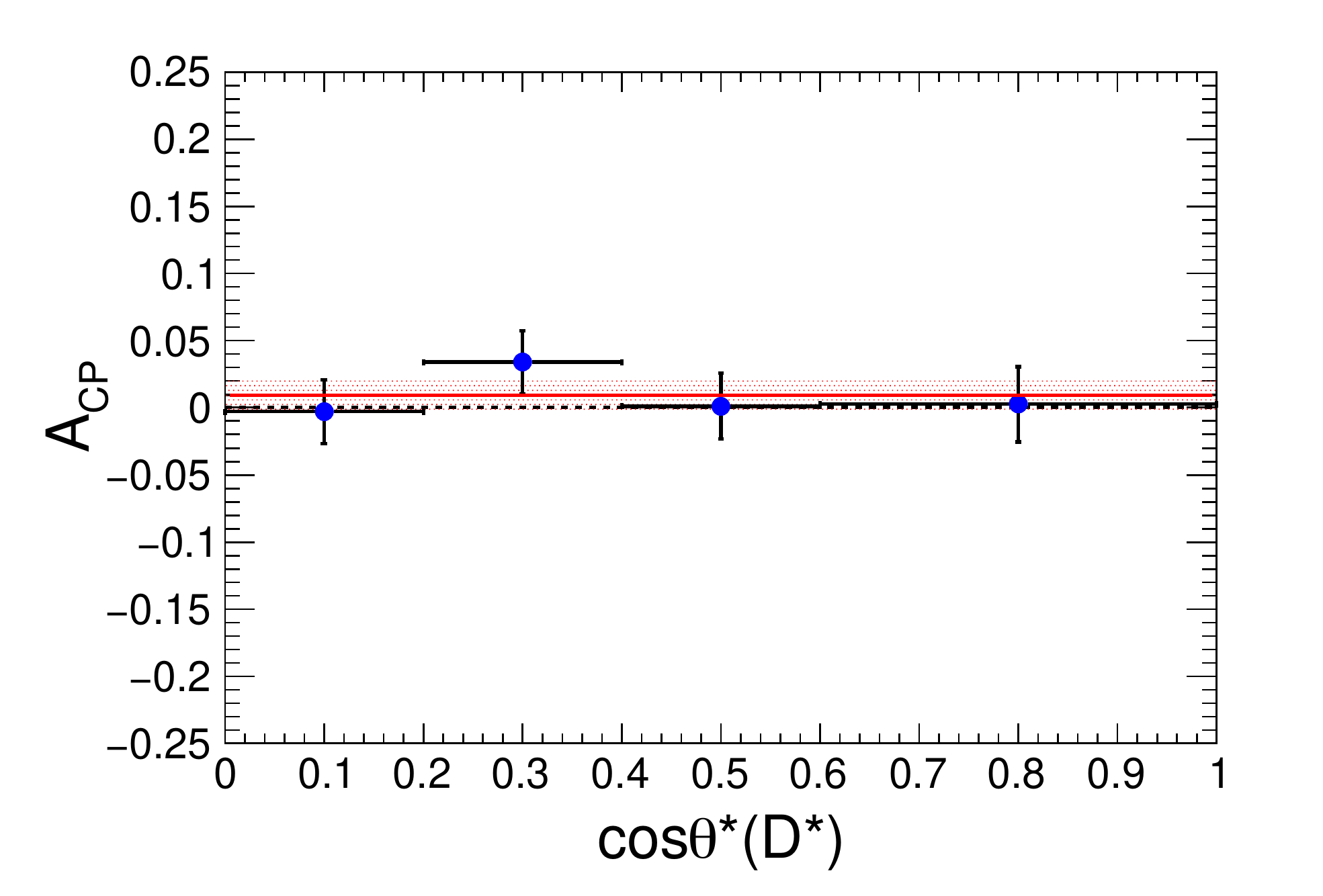}%
  \put(25,55){(a)}
  \end{overpic}%
  \begin{overpic}[width=0.333\textwidth]{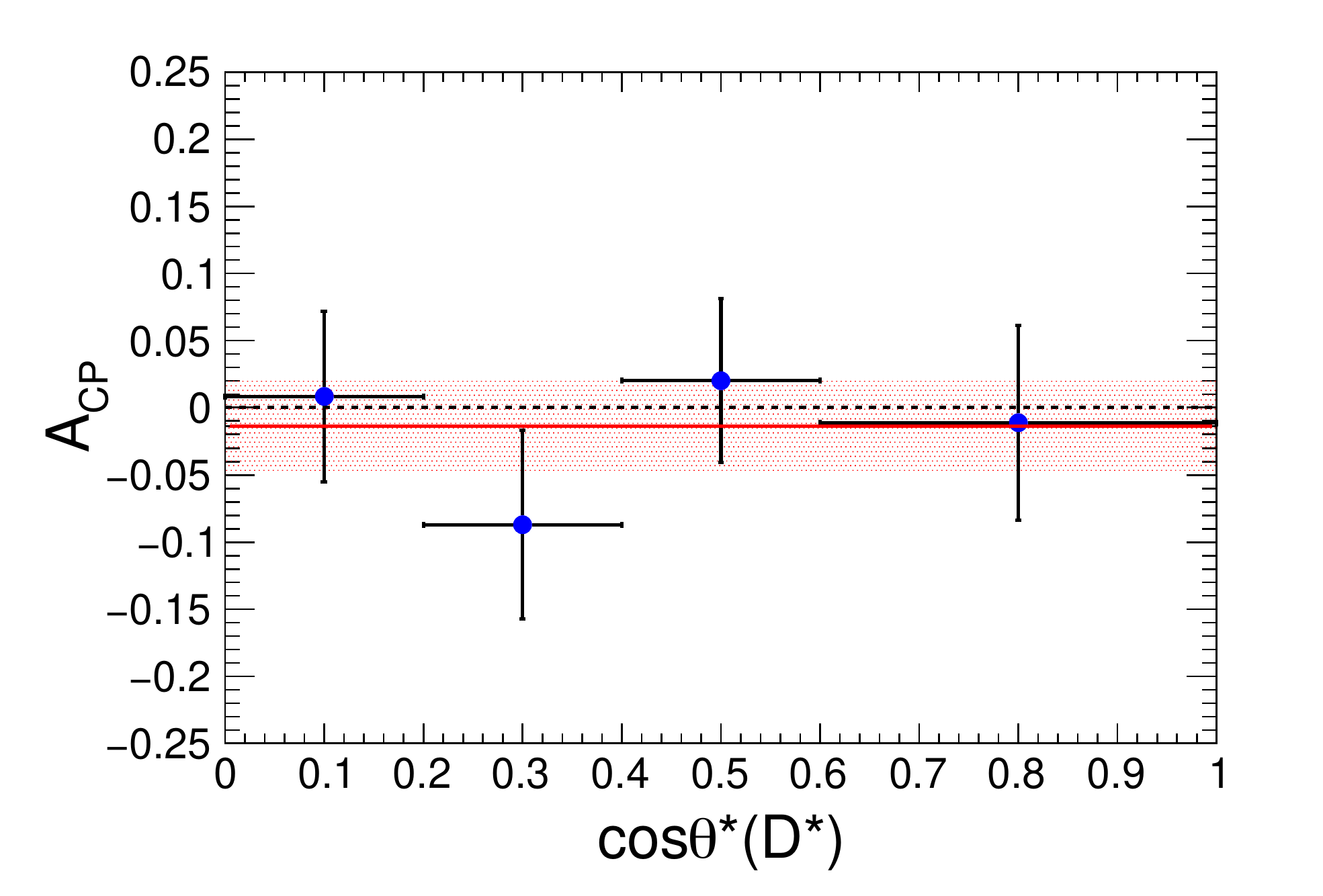}%
  \put(25,55){(b)}
  \end{overpic}%
  \begin{overpic}[width=0.333\textwidth]{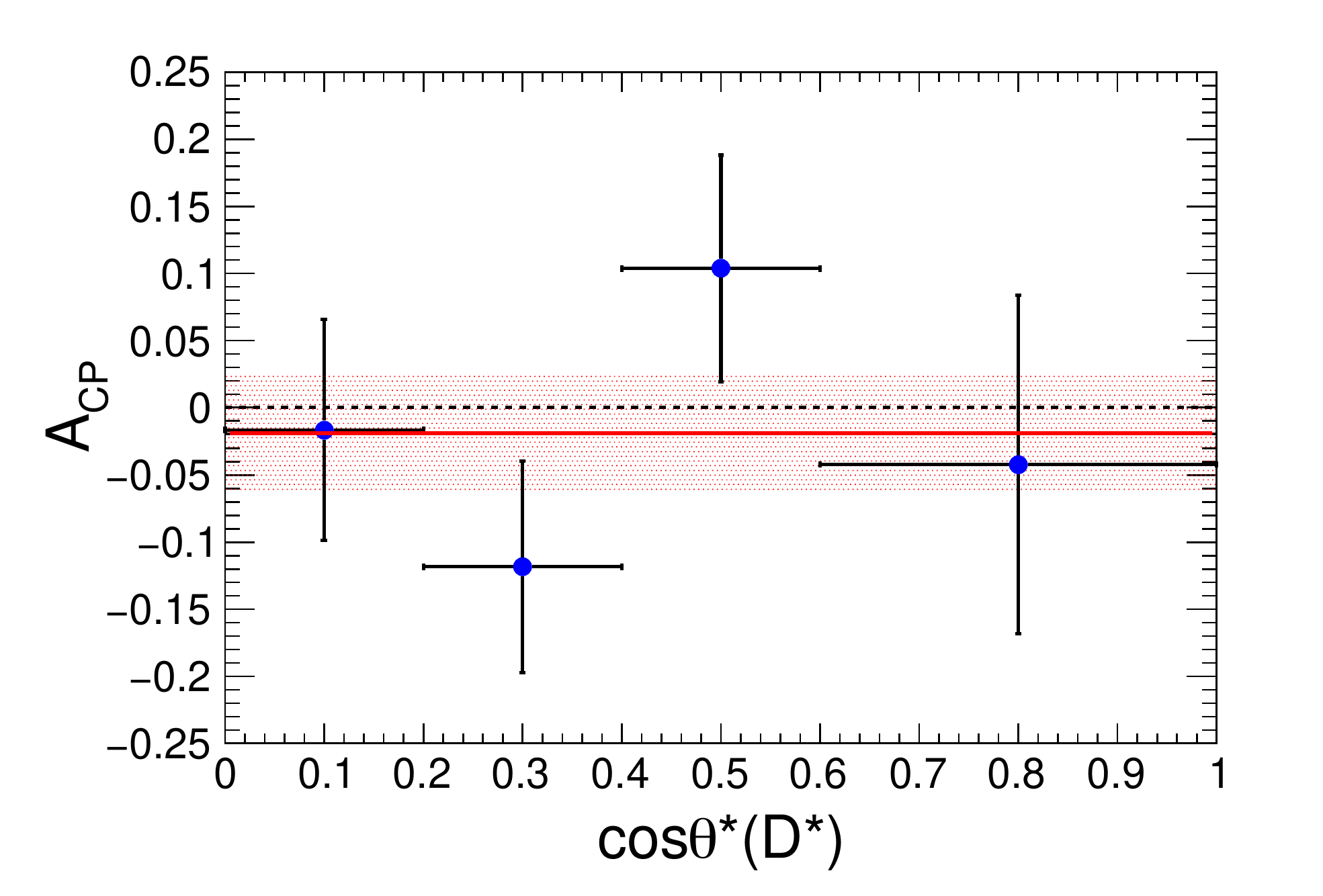}%
  \put(25,55){(c)}
  \end{overpic} 
  \vskip-5pt
  \caption{$\CP$-violating asymmetry $\Acp$ values as a function of $\cos\theta^{*}(D^{*+})$ for (a) $\DzToPiPiEta$, (b) $\DzToKKEta$, and (c) $\DzToPhiEta$, respectively. The solid red lines with a band region are the averaged values with their uncertainties. }
  \label{fig:AcpDzTohhEta}
  \end{center}
\end{figure}

We divide the $\Dz$ and $\Dzb$ samples into eight bins of $\cos\theta^{*}$. The events are weighted according to slow pion detection asymmetry map. 
We perform $Q$ or $M_{KK}$-$Q$ fit for weighted $D^0$ and $\Dzb$ samples simultaneously in each $\cos\theta^{*}$ bin, to extract the signal yield and corrected raw asymmetry.
Thus, using the formula Eq.(\ref{eqn:Acp}), we have four values of $\CP$ asymmetries and forward-backward asymmetries.
Fitting these $\Acp$ values in Fig.~\ref{fig:AcpDzTohhEta} to a constant gives us the final $\Acp$ results: 
\begin{eqnarray}
A_{\CP}(\DzToPiPiEta) & = & [0.9\pm 1.2\,({\rm stat})\pm 0.5\,({\rm syst})]\%, \\
A_{\CP}(\DzToKKEta) & = & [-1.4\pm 3.3\,({\rm stat})\pm 1.1\,({\rm syst})]\%, \\
A_{\CP}(\Dz\to\phi\eta) & = & [-1.9\pm 4.4\,({\rm stat})\pm 0.6\,({\rm syst})]\%.
\end{eqnarray}
The first result represents a significant improvement in precision over previous result~\cite{bib:PRD101d052009}. The latter two results are first measurements results. In these decays, no evidence for $\CP$ violation is found.

\subsection{$\Acp$ and $\mathcal{B}$ for $D_s^+\to\Kp\piz,\,\Kp\eta$ and $D_s^+\to\pip\piz,\,\pip\eta$ decays}
We measured the branching fractions and $\CP$ asymmetries for $D_s^+\to\Kp\piz,\,\Kp\eta$ and $D_s^+\to\pip\piz,\,\pip\eta$ decays~\cite{Belle:2021ygw}. 
Both $D_s^{*+}$-tagged and untagged $\Dsp$ samples from 921 $\invfb$ data set are used to measure the branching fractions relative to $D_s^+\to\phi\pip$ decay.
All the signal channels have only one charged track, except the channels with $\eta\to\pip\pim\piz$. Therefore, there is huge background. To suppress the background, we use a neural network. 
We perform a simultaneous fit on the tagged and untagged samples to extract signal yield. The results are shown in Tab.~\ref{tab:BRofDsToPP}.
Considering no significant signal for $D_s^+\to\pip\piz$ is observed, we set an upper limit for its branching fraction $\mathcal{B}(D_s^+\to\pip\piz)<1.2\times10^{-4}$ at $90\%$ conference level. 
\begin{table}[!htbp]
\begin{center}
\caption{\label{tab:BRofDsToPP}Reconstruction efficiencies, fitted signal yields, and resulting relative and absolute branching fractions. The yields listed are the sums of those from the tagged and untagged $\Dsp$ samples.}
\vskip-5pt
\begin{lrbox}{\tablebox}
\begin{tabular}{lcccc} \toprule 
Decay mode &  $\varepsilon$ (\%) & Fitted yield &    
$\mathcal{B}/\mathcal{B}_{\phi\pi^+}$ (\%)  & $\mathcal{B}$ ($10^{-3}$)  \\
\midrule 
$D_s^+ \rightarrow K^+ \pi^0$  & 
$8.10 \pm 0.04$  	& $11978 \pm 846$   	&  $3.28\pm 0.23\pm 0.13$  	& $0.735\pm 0.052\pm 0.030\pm 0.026$   \\
$D_s^+ \rightarrow K^+ \eta_{\gamma\gamma}$ & 
$7.42 \pm 0.05$  	& $10716 \pm 429$   	&  $8.04\pm 0.32\pm 0.35$  	& $1.80\pm 0.07\pm 0.08\pm 0.06$  \\
$D_s^+ \rightarrow K^+ \eta_{3\pi}$  & 
$4.04 \pm 0.02$  	& $3175 \pm  121$   		&  $7.62\pm 0.29\pm 0.33$  	& $1.71\pm 0.07\pm 0.08\pm 0.06$     \\
$D_s^+ \rightarrow K^+ \eta$   &
$-$ 				&  $-$  				& $7.81\pm 0.22\pm 0.24$ 	& $1.75\pm 0.05\pm 0.05\pm 0.06$  \\ \hline
$D_s^+ \rightarrow \pi^+ \pi^0$ & 
$6.63 \pm 0.04$  	&  $491 \pm 734$    		&  $0.16\pm 0.25\pm 0.09$   	& $0.037\pm 0.055\pm 0.021\pm 0.001$   \\
$D_s^+ \rightarrow \pi^+ \eta_{\gamma\gamma}$ & 
$10.84 \pm 0.02$ 	& $166696 \pm 1173$  	& $85.54\pm 0.64\pm 3.32$  	& $19.16\pm 0.14\pm 0.74\pm 0.68$ \\
$D_s^+ \rightarrow \pi^+ \eta_{3\pi}$ & 
$6.50 \pm 0.03$     	&  $56132 \pm 407$  	&  $83.55\pm 0.64\pm 4.37$  	& $18.72\pm 0.14\pm 0.98\pm 0.67$ \\
$D_s^+ \rightarrow \pi^+ \eta$ & 
$-$ 				& 	$-$ 				& $84.80\pm 0.47\pm 2.64$ 	& $19.00\pm 0.10\pm 0.59\pm 0.68$  \\
\hline
 $D_s^+ \rightarrow \phi \pi^+$ & 
 $22.05 \pm 0.13$  	&  $1005688\pm 2527$  	&   1  					& $-$  				\\
\bottomrule
\end{tabular}
  \end{lrbox}
  \scalebox{0.9}{\usebox{\tablebox}}
\end{center}
\end{table}

A simultaneous fit on four $M(D_s^+)$ distributions from $D_s^+$ and $D_s^-$ tagged and untagged samples to extract the raw asymmetry.
For $D_s^+\to\pip\eta$, we use reference mode $D\to\phi\pip$ which has already been well-measured.
Firstly we obtain the raw asymmetry difference between signal modes and reference modes to cancel the common asymmetry sources.
Then, by adding the reference mode’s $\CP$ asymmetry which has been measured before, we obtain the $\Acp$ for $D_s^+\to\pip\eta$.
For $D_s^+\to\Kp\piz$ and $\Kp\eta$, there is no suitable reference mode, we firstly perform simultaneous fit for six bins of $\cos\theta^{*}(D_s^+)$, separately, to obtain the raw asymmetries; and add a correction based on kaon detection asymmetry map. Then we simultaneously fit the weighted events to obtain the corrected raw asymmetry in six bins of $\cos\theta^{*}(D_s^+)$, separately. 
Finally, fitting these $\Acp$ to a constant gives our final $\CP$ asymmetries in Tab.~\ref{tab:AcpDsToPP}.
These results show no evidence of $\CP$ violation.
\begin{table}[!htbp]
\begin{center}
\caption{\label{tab:AcpDsToPP}Measured $\CP$ asymmetries. The first and second uncertainties listed are statistical and systematic, respectively. Results from the two $\eta$ decay modes are combined via a weighted average and also listed.}
\vskip-5pt
\begin{tabular}{lccc}
\toprule 
Decay mode &   $A_{\rm {raw}}$  & $A_{\CP}$  \\ \midrule 
$D_s^+ \rightarrow K^+ \pi^0$          	&   \phantom{$-$}0.115 $\pm$ 0.045   	&  \phantom{$-$}0.064  $\pm$ 0.044 $\pm$ 0.011       \\
$D_s^+ \rightarrow K^+ \eta_{\gamma\gamma}$ 
							&   \phantom{$-$}0.046 $\pm$ 0.027   	&  \phantom{$-$}0.040   $\pm$ 0.027 $\pm$ 0.005    \\
$D_s^+ \rightarrow K^+ \eta_{3\pi}$  	&   $-$0.011 $\pm$ 0.033 				&  $-$0.008  $\pm$ 0.034 $\pm$ 0.008      \\
$D_s^+ \rightarrow K^+ \eta$           	&   \phantom{$-$}$-$ 				& \phantom{$-$}0.021 $\pm$ 0.021 $\pm$ 0.004    \\
\hline
$D_s^+ \rightarrow \pi^+ \eta_{\gamma\gamma}$   
							& \phantom{$-$}0.007 $\pm$ 0.004   	&   \phantom{$-$}0.002 $\pm$ 0.004 $\pm$ 0.003  \\
$D_s^+ \rightarrow \pi^+ \eta_{3\pi}$	&   \phantom{$-$}0.008 $\pm$ 0.006    	&  \phantom{$-$}0.002 $\pm$ 0.006 $\pm$ 0.003  \\
$D_s^+ \rightarrow \pi^+ \eta$          &  \phantom{$-$}$-$ 					& \phantom{$-$}0.002 $\pm$ 0.003 $\pm$ 0.003   \\
\hline
 $D_s^+ \rightarrow \phi\pi^+$           & \phantom{$-$}0.002 $\pm$ 0.001 		& $-$    \\
\bottomrule 
\end{tabular}
\end{center}
\end{table}

\section{Branching fractions for the decays of charmed baryons}
\subsection{$\mathcal{B}$ of $\Lambda_c^+\to p\piz$ and $p\eta$}
We perform a measurement of branching fractions for SCS decays $\Lambda_c^+\to p\piz$ and $p\eta$~\cite{Belle:2021mvw}.
These decays proceed predominantly through internal W emission and W exchange. The theoretical calculations predict $\mathcal{B}(\Lcp\to p\eta)$ to be at least one order of magnitude greater than $\mathcal{B}(\Lcp\to p\piz)$.
BESIII reported the first evidence of $\Lambda_c^+\to p\eta$; and set an upper limit for branching fraction for $\Lambda_c^+\to p\piz$ at 90\% C.L.
We use the 980 $\invfb$ data set at Belle to measure the branching fractions relative to reference CF mode $\Lambda_c^+\to p\Km\pip$. 
We obtain about 8 thousand $\Lambda_c^+\to p\eta$ signals, and no significant $\Lambda_c^+\to p\piz$ signal, 
A large sample for reference mode and its averaged efficiency from MC which is adjusted on Dalitz plot.
Finally, we obtain the relative branching fractions.
$\frac{\mathcal{B}(\Lambda_c^+\to p\eta)}{\mathcal{B}(\Lambda_c^+\to p\Km\pip)}=[2.258\pm 0.077({\rm stat})\pm 0.122({\rm syst})]\times 10^{-2}$, and $\frac{\mathcal{B}(\Lambda_c^+\to p\piz)}{\mathcal{B}(\Lambda_c^+\to p\Km\pip)}<1.273\times 10^{-3}$ at 90\% C.L.
Using world average $\mathcal{B}(\Lcp\to p\Km\pip)$~\cite{bib:PDG2021}, we have $\mathcal{B}(\Lambda_c^+\to p\eta)=[1.42\pm 0.05({\rm stat})\pm 0.11({\rm syst})]\times 10^{-3}$, consistent with and more precise than other results.
An upper limit on BF of $\Lambda_c^+\to p\piz$ is set: $\mathcal{B}(\Lambda_c^+\to p\piz)<8.0\times 10^{-5}$.
The ratio of these two branching fractions is consistent with the theoretical prediction.

\subsection{$\mathcal{B}$ of $\Lambda_c^+\to\Lambda\pip\eta$, $\Sigma^0\pip\eta$, $\Lambda(1670)\pip$, and $\eta\Sigma(1385)$}
\begin{figure}[!htbp]
  \begin{center}
     \begin{overpic}[width=0.33\textwidth,height=0.24\textwidth]{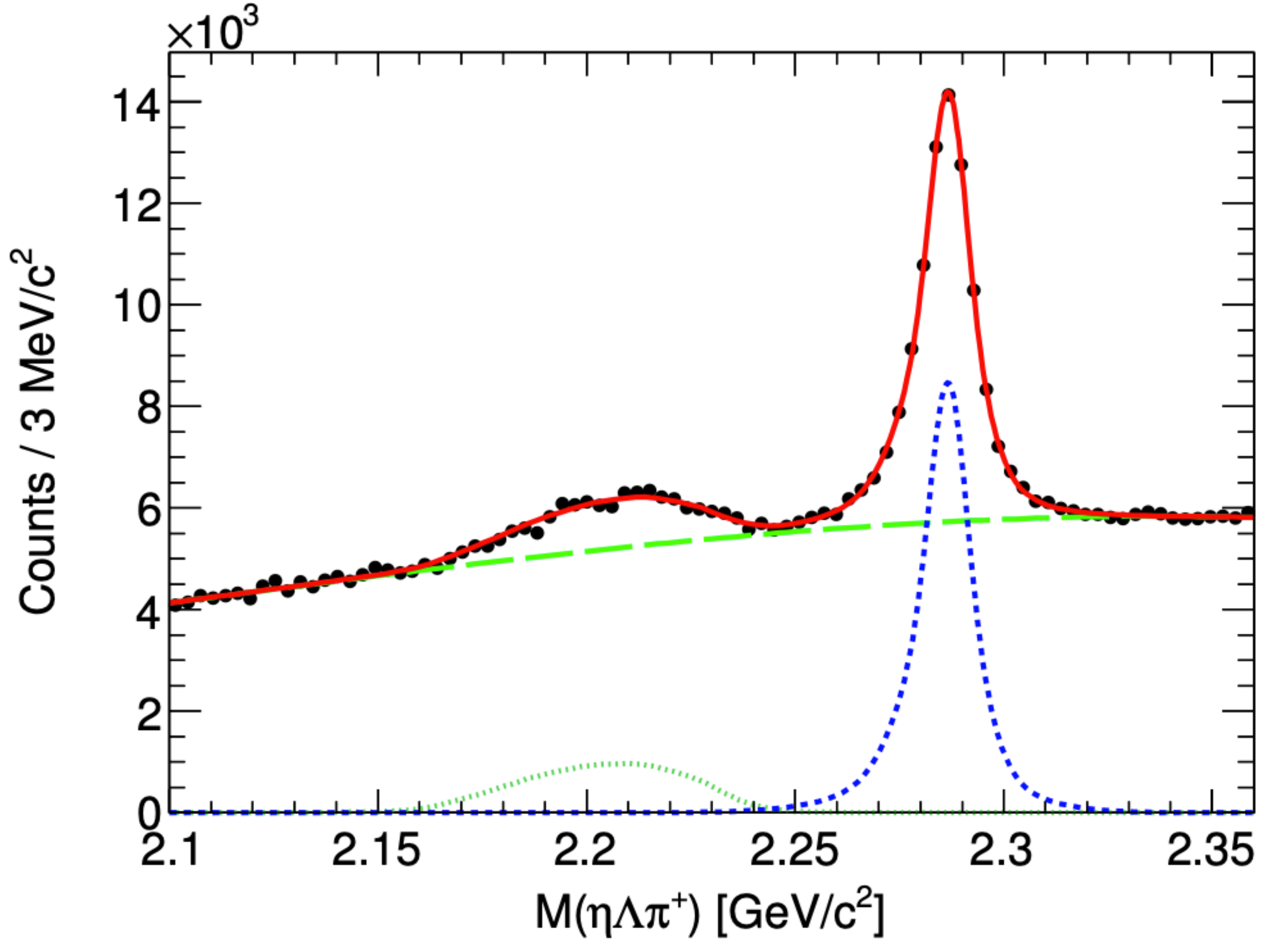}%
     \end{overpic}%
     \begin{overpic}[width=0.33\textwidth,height=0.24\textwidth]{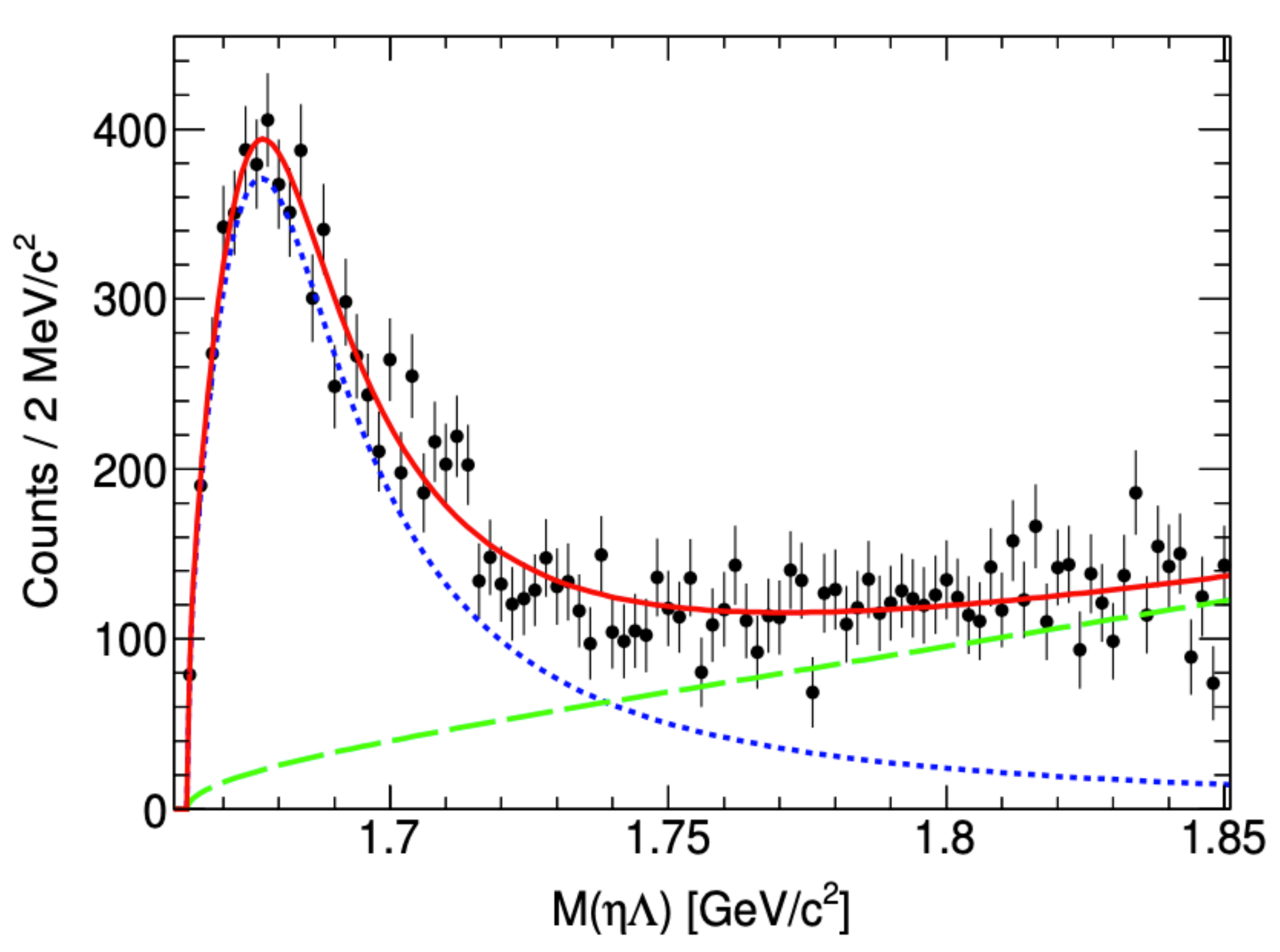}%
     \end{overpic}%
     \begin{overpic}[width=0.33\textwidth,height=0.24\textwidth]{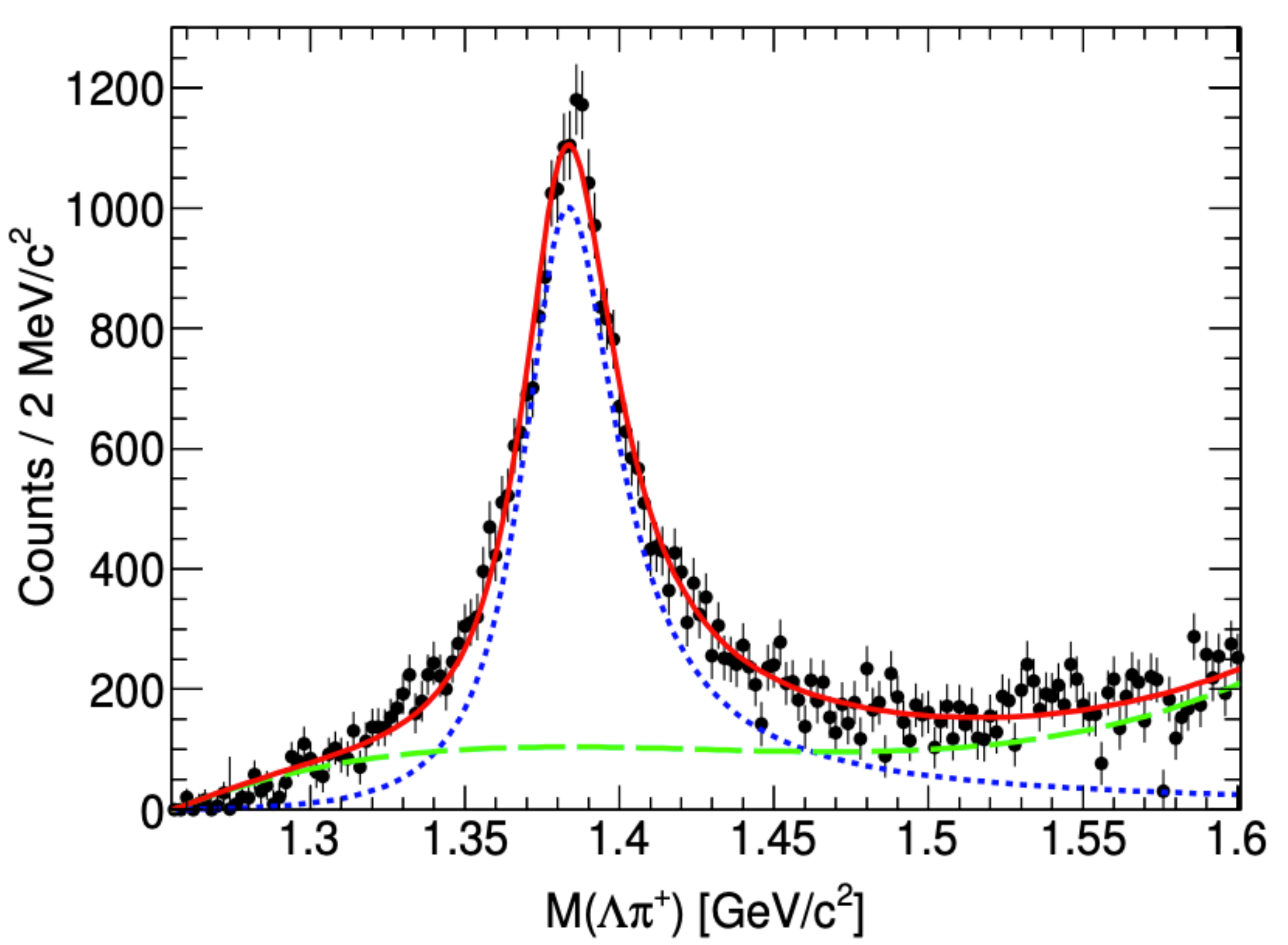}%
     \end{overpic}%
     \vskip-5pt
   \caption{Left figure is the invariant mass of $\eta\Lambda\pip$; middle figure is the invariant mass of $\eta\Lambda$ to extract $\Lambda(1670)$ info; right figure is the invariant mass of $\Lambda\pip$ to extract $\Sigma^+(1385)$ info.  }
  \label{fig:LcToHV}
  \end{center}
\end{figure}

Here we introduce a study on $\Lambda_c^+$ multi-body decay: $\Lambda_c^+\to\Lambda\pip\eta$ decay~\cite{Belle:2020xku}. 
This decay is an ideal mode to study $\Lambda(1670)$ because the isospin is fixed for any combination of two particles in final state.
In the $\Lambda_c^+$ invariant mass distribution in Fig.~\ref{fig:LcToHV} (left), we obtain a large sample of $\Lambda_c^+\to\Lambda\eta\pip$. Meanwhile, as a feed-down component, $\Lambda_c^+\to\Sigma^0\eta\pip$ is obtained. 
For later channel, an average efficiency of three-body decay is used to measure its branching fraction. 
For former one and the reference mode $\Lambda_c^+\to p\Km\pip$, they both have sufficiently large statistics to extract yield in individual bins of Daltiz plots. This is to consider the bin-to-bin variations of efficiencies. 
Thus, we have the relative branching fraction, \begin{eqnarray}
\mathcal{B}(\Lambda_c^+\to\eta\Lambda\pip)/\mathcal{B}(\Lambda_c^+\to p\Km\pip) & = & 0.293\pm0.003\pm0.014\,, \\
\mathcal{B}(\mathcal{B}(\Lambda_c^+\to\eta\Sigma^0\pip)/\mathcal{B}(\Lambda_c^+\to p\Km\pip) & = & 0.120\pm0.006\pm0.010\,.
\end{eqnarray}
Finally, using world average $\mathcal{B}(\Lcp\to p\Km\pip)$~\cite{bib:PDG2021}, we obtain the absolute branching fractions: $\mathcal{B}(\Lambda_c^+\to\eta\Lambda\pip) = (1.84\pm 0.02\pm 0.09 \pm 0.09)\%$ and $\mathcal{B}(\Lambda_c^+\to\eta\Sigma^0\pip) = (7.56\pm 0.39\pm 0.62 \pm 0.39)\times10^{-3}$.

On the Dalitz plot of $\Lambda_c^+\to\Lambda\eta\pip$, we find three bands corresponding to $\Lambda(1670)$, $\Sigma(1385)$, and $a_0(980)$. 
We extract the $\Lambda_c^+$ signal yield in every 2 MeV bin of the $\eta\Lambda$ and $\Lambda\pi$ invariant mass in Fig.~\ref{fig:LcToHV}. Then a relativistic Breit-Wigner is used to describe S-wave $\Lambda(1670)$ and P-wave $\Sigma(1385)$. 
The branching fractions of these two intermediate processes are measured, 
\begin{eqnarray}
\mathcal{B}(\Lambda_c^+\to[\Lambda(1670)\to\eta\Lambda]\pip)/\mathcal{B}(\Lambda_c^+\to p\Km\pip) & = & (5.54\pm0.29\pm0.73)\times10^{-2}\,, \\
\mathcal{B}(\Lambda_c^+\to\eta\Sigma(1385)^+)/\mathcal{B}(\Lambda_c^+\to p\Km\pip) & = & 0.192\pm 0.006\pm 0.016\,,
\end{eqnarray}
along with their fitted mass and width values:
$M=1674.3\pm0.8\pm4.9$ MeV/$c^2$ and $\Gamma=36.1\pm2.4\pm4.8$ MeV for $\Lambda(1670)$; 
$M=1384.8\pm0.3\pm1.4$ MeV/$c^2$ and $\Gamma=38.1\pm1.5\pm2.1$ MeV for $\Sigma(1385)$.

\subsection{$\mathcal{B}$ and $\alpha$ of $\Xi_c^0\to\Lambda\Kstarzb,\,\Sigma^0\Kstarzb$ and $\Sigma^+\Kstarm$}
We measure the branching fractions and asymmetry parameters for $\Xi_c^0\to\Lambda\Kstarzb,\,\Sigma^0\Kstarzb$ and $\Sigma^+ \Kstarm$~\cite{Belle:2021zsy}. The motivation is that the branching fraction measurements help to distinguish different theoretical modes; the asymmetry parameter of $\Xi_c$ decays are still not well measured, which is important to test parity violation in charmed baryons sectors.
The branching fractions are measured normalizing to reference mode $\Xi_c^0\to\Xi^-\pip$, and based on the data set of 980 $\invfb$.
The signal yields are extracted via $M(\Xi_c^0)$-$M(K^{*})$ two-dimensional fit.
Then considering the efficiency correction, we have the relative branching fractions of $\Xi_c^0\to\Lambda\Kstarzb$, $\Xi_c^0\to\Sigma^0\Kstarzb$ and $\Xi_c^0\to\Sigma^{+}K^{*-}$ relative to $\Xi_c^0\to\Xi^-\pip$: $0.18\pm 0.02 \pm 0.01$, $0.69\pm 0.03 \pm 0.03$, and $0.34\pm 0.06 \pm 0.02$, respectively, where the errors are statistical and systematic uncertainties. 
This is first such measurement.

Taking $\Xi_c\to\Lambda\Kstarzb$ for example, the differential decay rate is $\frac{dN}{d\cos\theta_{\Lambda}} \propto  1+\alpha(\Xi_c^0\to\Lambda\Kstarzb)\alpha(\Lambda\to p\pim)\cos\theta_{\Lambda}$, which is relative to asymmetry parameter $\alpha$ of $\Xi_c^0$ and $\Lambda$ decays, and $\theta_{\Lambda}$, the $\Lambda$ helicity angle between the proton momentum and $\Kstarzb$ momentum in the $\Lambda$ rest frame.
As to the asymmetry parameter of ($\Sigma^0\to\gamma\Lambda$) is zero for an electromagnetic decay. So, we cannot measure the asymmetry parameter of $\Xi_c^0\to\Sigma^0\Kstarzb$ only based on $\theta_{\Sigma^0}$. But we measure their product value just to validate no bias in the measurement.

The yields with the efficiency-correction dependent on the cosine of helicity angle are fitted with a linear function to extract asymmetry parameters. This fitted slope for $\Xi_c^0\to\Sigma^0\Kstarzb$ is consistent with zero, which shows no bias. 
And using the differential decay rate formula and the asymmetry values for $\Lambda/\Sigma^+\to p\pi^{-,0}$~\cite{bib:PDG2021}, we finally have asymmetry parameters:
$\alpha(\Xi_c^0\to\Lambda\Kstarzb)=0.15\pm 0.22\pm 0.05$ and $\alpha(\Xi_c^0\to\Sigma^+\Kstarm)=-0.52\pm 0.30\pm 0.02$,
for the first time.

\section{Properties of excited charmed baryons}

\subsection{electromagnetic transition $\Xi_c(2790,\,2815)^{+,0}\to\Xi_c^{+,0}\gamma$}
Here we present a search for four new electromagnetic decays $\Xi_c(2790,\,2815)^{+,0}\to\Xi_c^{+,0}\gamma$~\cite{Belle:2020ozq}.
Firstly, we obtain a large sample of $\Xi_c^{0,+}$: 142 thousands $\Xi_c^0$ and 79 thousands $\Xi_c^+$, using ten or seven decay modes respectively.
Then we fit the $\Xi_c\gamma$ invariant mass distribution with a Breit-Wigner convoluted Crystal Ball for signal, and a polynomial for background. We achieve the first observation (evidence) of $\Xi_c(2815)^0$ ($\Xi_c(2790)^0$) electromagnetic decays, respectively. Their branching fractions relative to their responding reference modes are measured. 
\begin{eqnarray}
\frac{\mathcal{B}(\Xi_c(2815)^{0}\to\Xi_c^{0}\gamma)}{\mathcal{B}(\Xi_c(2850)^{0}\to\Xi_c(2645)^{+}\pim\to\Xi_c^{0}\pip\pim)} & = & 0.41\pm0.05\pm0.03\,, \\
\frac{\mathcal{B}(\Xi_c(2790)^{0}\to\Xi_c^{0}\gamma)}{\mathcal{B}(\Xi_c(2790)^{0}\to\Xi_c^{\prime{+}}\pim\to\Xi_c^{+}\gamma\pim)} & = & 0.13\pm0.03\pm0.02\,, 
\end{eqnarray}
We do not find evidence for the analogous decays for $\Xi_c(2815)^{+}$ and $\Xi_c(2790)^{+}$, an upper limit at 90\% C.L. are set respective to reference mode: $<0.09$ and $<0.06$, respectively.

\subsection{determination of $J^{P}(\Xi_c(2970)^+)$}
The first determination of spin-parity of a charmed-strange baryon, $\Xi_c(2970)^+$, is presented in Ref.~\cite{Belle:2020tom}.
The unclear theoretical situation motivates this experimental determination, which provides important information to test predictions and help decipher its nature. 

The spin of $\Xi_c(2970)^+$ is determined by angular analysis of this decay chain $\Xi_c(2970)^+\to\Xi_c(2645)^0\pip\to\Xi_c~+\pim\pip$: 
(1) helicity angle of $\Xi_c(2970)$: we fit the background-subtracted and efficiency-corrected yields with expected decay-angle distribution for different spin hypotheses. The best fit is for $J=1/2$ but this is inconclusive, since others hypotheses are excluded with small significance.
(2) helicity angle of $\Xi_c(2645)$: the expected angular correlation is used to fit. The half-spin hypothesis over the others at the level of $5.1\sigma$ and $4.0\sigma$, respectively.

The parity of $\Xi_c(2970)^+$ is established from the relative branching fraction: $\mathcal{B}(\Xi_c(2970)^+\to\Xi_c(2645)^0\pip)/\mathcal{B}(\Xi_c(2970)^+\to\Xi_c^{\prime0}\pip)=1.67\pm0.29(\text{stat})^{+0.15}_{-0.09}(\text{syst})\pm0.25(\text{IS})$, where the last uncertainty is due to possible isospin-symmetry-breaking effects. This result favors $J^P =\frac{1}{2}^{+}$ with the zero spin of the light-quark degrees of freedom.

\section{Summary}
Fruitful charm results are achieved at Belle experiment since last CHARM workshop. Some selected recent ones are presented here. 
More charm results from Belle are on the road. 
As a summary, I would like to say, our Belle is not only keeping alive but still keeping energetic with fruitful charm results, although its final full data set was achieved more than ten years ago.


\end{document}